\documentclass[twocolumn,showpacs,preprintnumbers,superscriptaddress]{revtex4-2}
\usepackage{float}
\usepackage{amsmath}
\usepackage{amsfonts}
\usepackage{dcolumn}
\usepackage{bm}
\usepackage{epsfig}
\usepackage{epstopdf}
\usepackage{graphicx}
\usepackage{amsfonts}
\usepackage{amssymb}
\usepackage{mathrsfs}
\usepackage{pifont}
\usepackage{color}
\usepackage{appendix}
\usepackage{multirow}
\usepackage{threeparttable}
\usepackage{verbatim}
\usepackage{graphicx,amsmath}
\usepackage{color,xcolor}
\usepackage[bookmarks=false]{hyperref}
\hypersetup{colorlinks=true,citecolor=blue,linkcolor=blue,urlcolor=blue,pdfstartview=FitH,bookmarksopen=true}

\begin{document}

\title{Time Crystal in the Nonlinear Phonon Mode of the Trapped Ions}

\author{Yi-Ling Zhan}
\affiliation{Key Laboratory of Materials Physics, Ministry of Education, School of Physics and Laboratory of Zhongyuan Light, Zhengzhou University, Zhengzhou 450001, China}
\author{Chun-Fu Liu}
\affiliation{Key Laboratory of Materials Physics, Ministry of Education, School of Physics and Laboratory of Zhongyuan Light, Zhengzhou University, Zhengzhou 450001, China}
\author{J.-T. Bu}
\affiliation{Key Laboratory of Materials Physics, Ministry of Education, School of Physics and Laboratory of Zhongyuan Light, Zhengzhou University, Zhengzhou 450001, China}
\author{K.-F Cui}
\email{cuikaifeng@zzu.edu.cn}
\affiliation{Key Laboratory of Materials Physics, Ministry of Education, School of Physics and Laboratory of Zhongyuan Light, Zhengzhou University, Zhengzhou 450001, China}
\author{S.-L. Su}
%	\email{slsu@zzu.edu.cn}
\affiliation{Key Laboratory of Materials Physics, Ministry of Education, School of Physics and Laboratory of Zhongyuan Light, Zhengzhou University, Zhengzhou 450001, China}
\affiliation{Institute of Quantum Materials and Physics, Henan Academy of Sciences, Zhengzhou 450046, China}
\author{L.-L. Yan}
\email{llyan@zzu.edu.cn}
\affiliation{Key Laboratory of Materials Physics, Ministry of Education, School of Physics and Laboratory of Zhongyuan Light, Zhengzhou University, Zhengzhou 450001, China}
\affiliation{Institute of Quantum Materials and Physics, Henan Academy of Sciences, Zhengzhou 450046, China}
\author{Gang Chen}
\email{chengang971@163.com}
\affiliation{Key Laboratory of Materials Physics, Ministry of Education, School of Physics and Laboratory of Zhongyuan Light, Zhengzhou University, Zhengzhou 450001, China}

\begin{abstract}
Time crystals constitute a novel phase of matter defined by the spontaneous breaking of time-translation symmetry. Here we present a scheme to realize a continuous-time crystal of the vibrational phonon in the normal mode of two coupled ultra-cold ions. By utilizing two addressable standing-wave lasers and adiabatic elimination method, we generate a controllable nonlinear phonon mode with the well-designed efficient linear gain and nonlinear damping. By controlling these parameters to satisfy the phase transition conditions of Hopf bifurcation and limit cycle phase, it behaves as a stable dissipative dynamics over timescales significantly longer than the oscillation period, indicating the emergence of discrete time-translation symmetry breaking in the phonon mode, i.e., a phonon time crystal. We further numerically simulate this phonon time crystal by using accessible experimental parameters and also demonstrate a robustness to the initial thermal state and thermalization of phonon mode, spin dephasing, and the control errors of Rabi frequencies. These results provide a practical scheme for observing a time crystal in a nonlinear phonon mode and will advance the research of time crystals.
\end{abstract}

\maketitle

\section{Introduction}

Time crystals, analogous to conventional spatial crystals, represent a many-body quantum state where the time-translation symmetry of ground state  spontaneously breaks. Proposed by Wilczek and Shapere~\cite{TimeCrystals1,TimeCrystals2,TimeCrystals3}, time crystals envision a system capable of maintaining stable periodic motion without external energy input. However, theoretical studies reveal the fundamental constraints: Due to continuous symmetry of classical systems, the ground state must remain static to prevent energy divergence, i.e., a principle termed as the "no-go" theorem~\cite{Impossible1, Impossible2, Impossible3}. Consequently, Wilczek's original model failed to satisfy thermodynamic equilibrium requirements~\cite{Impossible2}. 

To realize the time crystals on the actual physical system, the concept of time crystals needs to be more loose, and then the discrete time crystals (DTCs), where the system, periodic driven by the Floquet control, is proposed by Krzysztof~\cite{PhysRevA.91.033617}. By applying an external field with the period $T$, the system can exhibit subharmonic responses with integer periods $ nT  (n > 1)$~\cite{PhysRevLett.116.250401}, thus, manifesting discrete time-translation symmetry breaking~\cite{annurev:/content/journals/10.1146/annurev-conmatphys-031119-050658}. Notably, even with Hamiltonian perturbations, the sufficient Ising interactions still enable the robust subharmonic responses~\cite{PhysRevLett.118.030401}. Experimental breakthroughs occur in 2017 where the discrete time crystals have been observed in different quantum systems, such as the ion trap system~\cite{Zhang2017}, disordered nitrogen-vacancy centers system~\cite{Choi2017}, and superconducting qubits~\cite{PhysRevLett.119.010602}, and recently it is more accurately realized in the quantum processor~\cite{PhysRevLett.119.010602,Mi2022,doi:10.1126/sciadv.abm7652}.

Benefiting from massive endeavor, a theoretical framework more consistent with Wilczek's original vision~\cite{TimeCrystals2} is proposed as the continuous time crystals (CTCs), which exhibits spontaneous breaking of continuous time-translation symmetry under a time-independent Hamiltonian or within a steady-state dissipative environment, arising from intrinsic self-organized dynamics of the system itself~\cite{PhysRevLett.123.210602,WOS:001148179200002,WOS:000972820500004}. Unlike floquet driven DTCs, CTCs do not require external periodic driving, thus realizing a genuine form of temporal order that persists autonomously in time. As one important category of CTCs, dissipative time crystals, due to the practical advantages in quantum technology, spur an abundant of theoretical researches~\cite{PhysRevLett.121.035301,PhysRevLett.120.040404,PhysRevLett.122.015701,PhysRevLett.123.260401}, whereafter realized in a Bose-Einstein condensate coupled to an optical cavity~\cite{PhysRevLett.127.043602} and then demonstrated in the dissipative systems~\cite{doi:10.1126/science.abo3382,doi:10.1126/science.add2015}. These researches confirm the feasibility of circumventing the no-go theorem through the environmental coupling and dissipative dynamics to realize time crystal dynamics~\cite{PhysRevResearch.6.033185,PhysRevResearch.6.013130,RieraCampeny2020timecrystallinityin,PhysRevResearch.6.013130,PhysRevA.110.012208,PhysRevResearch.2.022002}.

Recently, nonlinear systems with limit cycles prove to be another powerful platform to achieve CTCs~\cite{PhysRevLett.130.150401,wu2024dissipative}. By introducing linear damping, linear gain and nonlinear damping, a feasible approach based on the nonlinear system is proposed in Ref.~\cite{TimeCrystals}, where the system exhibits a metastable state with stable oscillations during its time evolution, accompanied by a dissipative evolution process with a timescale far exceeding the oscillation period. If there is an experimental system with precisely controlled dissipation processes and accessible experimental parameters, the time crystals may be well experimentally observed.

As an ideal experimental platform, ion traps can enable precise controls and also exhibit sophisticated engineering characteristics of dissipative mechanisms~\cite{PhysRevLett.110.110502,PhysRevA.98.042310,Cole_2021,PhysRevLett.128.050603,PhysRevLett.131.043605,PhysRevLett.133.090402}. For examples, various of gain and loss processes on the spin states and phonon modes can be effectively produced by controlling the detuning and Rabi frequencies of different lasers~\cite{PhysRevLett.77.4728,PhysRevLett.128.050603,PhysRevLett.133.090402}, establishing ion traps as superb platforms for studying non-equilibrium~\cite{An2015,PhysRevLett.120.010601} and dissipative phenomena~\cite{PhysRevLett.128.050603,PhysRevLett.133.090402}. Besides, the precise single-ion addressing capability ensures the experimental accuracy and controllability of dissipative channels on different ions~\cite{PhysRevA.60.145}. Another significant advantage lies in the high-fidelity fluorescence detection~\cite{10.1063/5.0055999,Schindler_2013,Okada_2006,Zhou_Fei_2010,PhysRevLett.119.193602} and noise controls~\cite{Charles_Doret_2012}, which provide the robust data support for the observation of time crystal properties. 

In this work, we demonstrate the experimental feasibility of realizing the continuous time crystal in the trapped ion system. The rest of this paper is organized as follows: In Sec.~\ref{section2}, we choose the appropriate Hamiltonian and Lindblad operators to construct the required scheme for the realization of the time crystals in the trapped ion system. Subsequently, by using the adiabatic elimination method, we derive an effective Hamiltonian and effective Lindblad operators, which together form the effective Lindblad master equation of time crystals. In Sec.~\ref{section3}, we perform a systematic analysis of the practically experimental parameters to validate the consistency of the theoretical scheme and its experimental feasibility. In Sec.~\ref{section4}, we use the accessible parameters of trap ion system to numerically solve the Lindblad master equation from a truncated Fock space, and demonstrate the characteristic behavior of time crystals: The system rapidly enters a metastable state with stable oscillations and subsequently evolves toward a steady state with a decay time significantly longer than the oscillation period. Then, we discuss the influences of initial  thermal state and thermalization of phonon mode, spin dephasing and control errors on time crystals. Finally, we give a brief conclusion in Sec.~\ref{section6}.

\section{Model of the System}
\label{section2}

\begin{figure}[htbp]
    \centering
    \includegraphics[width=8.7cm,height=2.8cm]{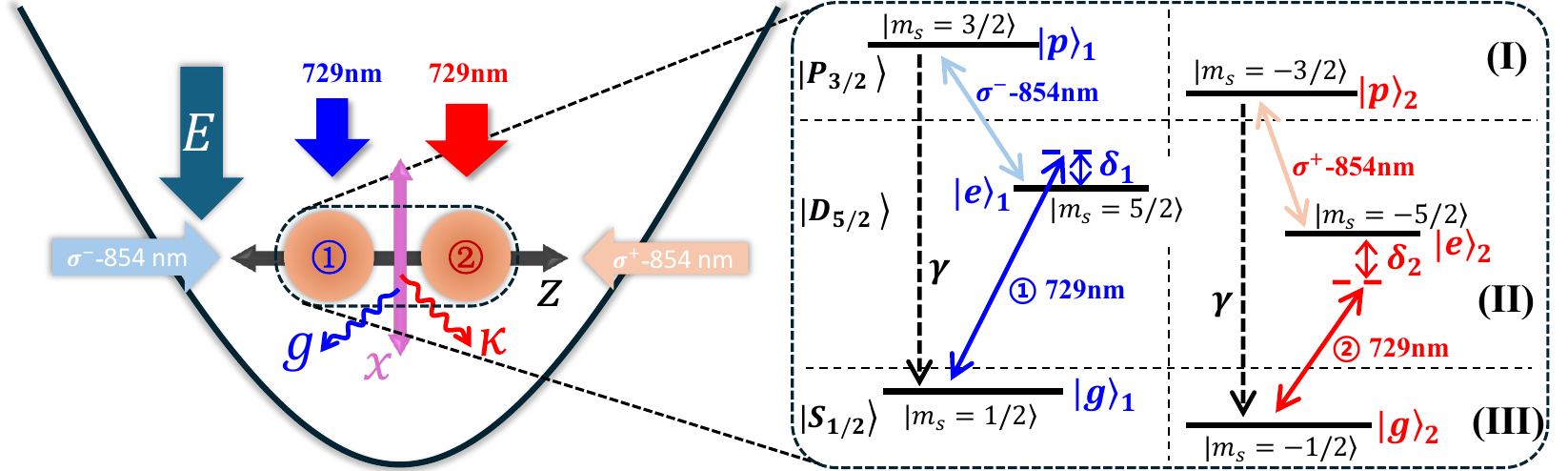}
    \caption{The scheme to realize the time crystal in the trapped $^{40}\mathrm{Ca}^{+}$ ion system, where an external alternating electric field along the radial direction ($x$-direction) of ions is applied to drive the normal vibrational mode of two ions, two polarized 854~nm lasers couple the metastable state $|e\rangle$ to the excited state $|p\rangle$ with the decay rate $\gamma$ to produce an effective dissipation process from the metastable state to ground state, and two standing wave laser beams of 729~nm with the detuning $\delta_{1,2}$ couple the assistant two-level system, consisting of the states $|g\rangle$ and $|e\rangle$, to the normal phonon mode of two ions to create the linear gain and nonlinear damping on the vibrational mode, respectively.}
    \label{fig1}
\end{figure}

As shown in Fig.~\ref{fig1}, we consider a system composed of one $^{40}\mathrm{Ca}^{+}$ ion trapped in the linear Paul trap, where the ion has two three-level structures as: the ground state $|g\rangle_{1,2} \equiv |4^2S_{1/2},m_s=\pm 1/2\rangle \notag$, metastable state $|e\rangle_{1,2} \equiv |3^2D_{5/2},m_s=\pm 5/2\rangle$ and excited state $|p\rangle_{1,2} \equiv |4^2P_{3/2},m_s=\pm 3/2\rangle $, respectively, and the radial vibrational mode of the ion serves as the phonon mode, which is driven by an external electric field with the amplitude $E$ and frequency $\omega_e$. To realize the time crystal dynamics in phonon mode of ions, we introduce the following laser fields: two 854~nm travelling lasers with $\sigma^{\mp}-$ polarization and two standing 729~nm lasers with different frequencies together consist of two three-level systems in these two ions, respectively, which are used to construct the desired effective linear gain and nonlinear damping dissipation channels on the phonon mode, respectively. In the following, we elaborate the constructed scheme by three steps (see Fig.~\ref{fig1}): In the step (I), we eliminate the excited state to produce an effective decay from the metastable state $|e\rangle$ to ground state; In the step (II), we eliminate the metastable state to produce the effective linear gain and nonlinear damping on the phonon mode; In the step (III), we trace over the spin state and add the driving term to realize the dissipative dynamics of phonon mode. 

First, we need to produce a large effective decay rate from the metastable state $|e\rangle$ to the ground state $|g\rangle$. However, due to the metastable state $|e\rangle$ owning a much long lifetime (about $1$~s), we utilize an auxiliary excited state $|p\rangle$ to accelerate this dissipative process. To realize this goal, we apply the polarized 854~nm laser to couple the metastable state $|e\rangle$ of ion to its excited state $|p\rangle$, producing the Hamiltonian as  ($\hbar = 1$)
\begin{equation}\label{Eq1}
H_{1} = \sum_{n=1,2} \left[ \frac{\omega_{e,n}}{2} \tilde{\sigma}_{z,n}+\frac{\Omega_{e,n}}{2}\left( \tilde{\sigma}_{-,n} e^{i\tilde{\omega}_{L,n} t} + h.c. \right)\right] ,
\end{equation}
where $n=1,2$ denote the  $n$th ion, $\omega_{e,n}$ is the energy difference between the metastable state $|e\rangle_n$ and excited state $|p\rangle_n$, $\Omega_{e,n}$ and $\tilde{\omega}_{L,n}$ are the Rabi frequency and frequency of $n$th 854nm laser, and the Pauli operators of $n$th ion are denoted as $\tilde{\sigma}_{-,n} = |e\rangle_n \langle p|$ and $\tilde{\sigma}_{z,n} = |p\rangle_n \langle p|- |e\rangle_n \langle e|$. To obtain an effective dissipative process from $|e\rangle_n$ to $|g\rangle_n$, we adiabatic eliminate the excited state by using the effective operator formalism method proposed in Ref.~\cite{EffH}. To obtain an effective dynamics and remove Stark shift induced by the 854~nm laser, we make the parameters to satisfy the conditions: 1) $\gamma \gg \Omega_{e,n}$ with $\gamma$ denoting the decay rate of excited state and 2) the detuning of 854~nm laser $\tilde{\delta}_n=\omega_{e,n}-\tilde{\omega}_{L,n}=0$, respectively. Then, the effective decay process from the metastable state $|e\rangle_n$ to ground state $|g\rangle_n$ is described as (see Appendix.~\ref{PA}) 
\begin{equation}
L_{\mathrm{eff}} = \sum_{n = 1}^2 i \sqrt{\Gamma_n} \sigma_{-,n}, \quad \Gamma_n = \frac{\Omega_{e,n}^2}{\gamma},
\label{Eq2}
\end{equation}
with the restriction $\gamma\gg\Omega_{e,n}\gg \Gamma_n$. 

\begin{figure}[htbp]
    \centering
    \includegraphics[width=8.5cm,height=5.2cm]{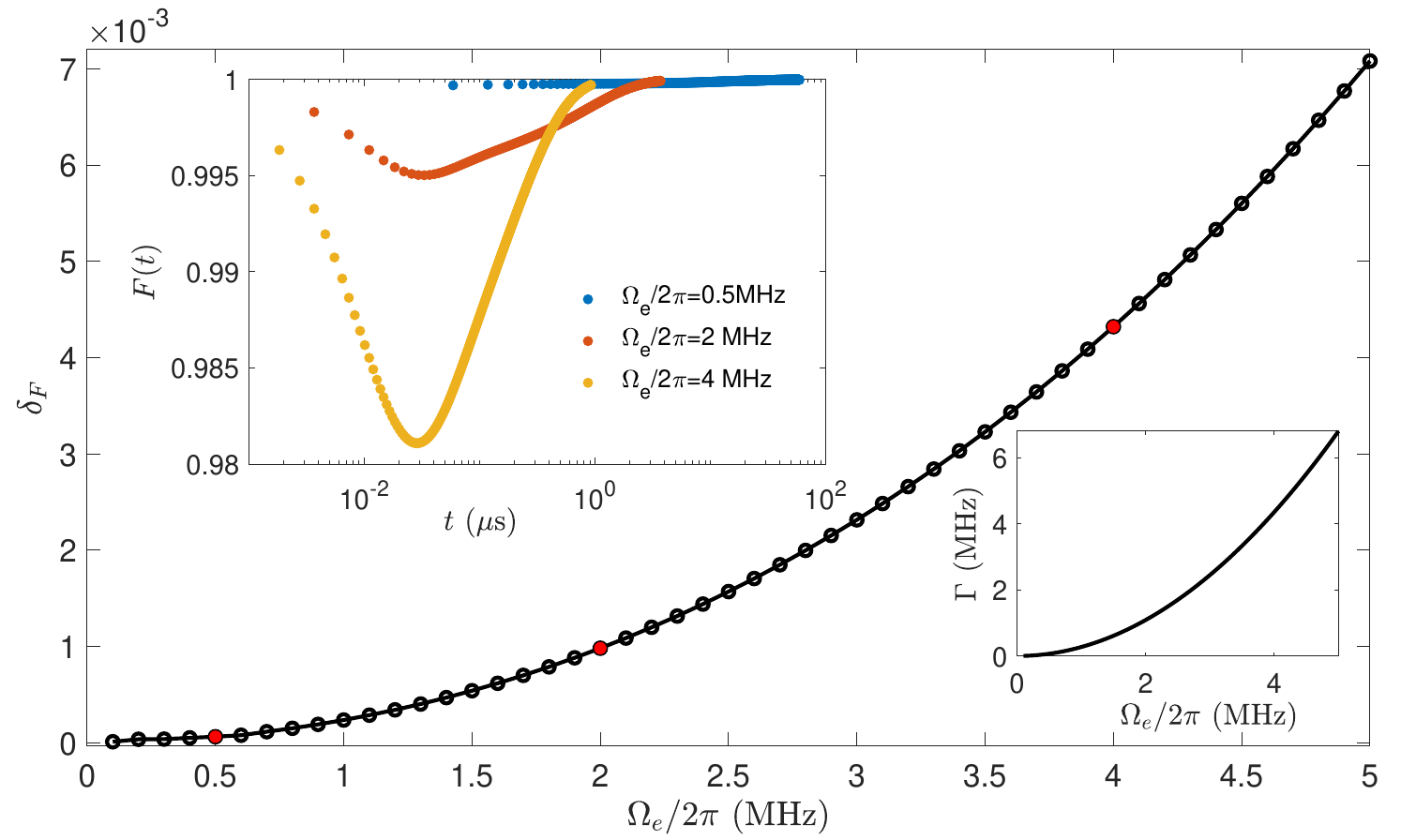}
    \caption{The  mean deviation $\delta_F$ between the numerical result and the effective dynamics, where the initial state of system is in the metastable state $|e\rangle$ and the insets denote the evolution of fidelity when $\Omega_e/2\pi=[0.5, 2, 4]$~MHz, respectively, and the effective dissipative rate $\Gamma$. Here, the decay rate of excited state $|p\rangle$ is chosen as $\gamma/2\pi=22.4$~MHz for the $^{40}\mathrm{Ca}^{+}$ ion, and the effective decay rate is given by $\Gamma = \Omega_e^2/\gamma$. }
    \label{figs1}
\end{figure}

To further characterize the validity of effective dissipative process, we use a mean deviation of fidelity $F(t)$ between the original dynamics and effective dynamics as $\delta_F=1-\tau^{-1}\int_0^{\tau}F(t)dt$, where $F(t)=\text{Tr}[\sqrt{\sqrt{\tilde{\rho}(t)}\rho(t)\sqrt{\tilde{\rho}(t)}}]$ with the evolution density matrix $\rho(t)$ of effective dynamics and $\tilde{\rho}(t)$ of original dynamics, respectively. In the Fig.~\ref{figs1}, it shows that the mean deviation $\delta_F$ rapidly increases as the Rabi frequency $\Omega_e$, indicating that the effective dissipative process deviates from the original dynamics and thus the adiabatic elimination becomes invalid for a large $\Omega_e$. Therefore, to obtain an effective dynamics, we generally restrict $\Omega_e/2\pi<2$~MHz, i.e., $\Omega_e/\gamma<0.1$, which gives the effective decay rate $\Gamma<1$~MHz (i.e., $\Gamma/2\pi<0.17$~MHz).

Then, we produce the linear gain ($g$) and nonlinear damping ($\kappa$) on the phonon mode by using two standing wave 729~nm lasers and adiabatically eliminating the metastable state $|e\rangle$~\cite{PhysRevLett.77.4728,EffH}. In our scheme, the ion is first cooled down to the ground state after the ground cooling and then illuminated by the two standing wave 729~nm lasers with the interaction Hamiltonian between the standing wave 729~nm lasers and atoms simplified as:
%with the beam waist smaller than the distance of two ions is separated, meanwhile, the ions are , making the wave packet of ions much smaller than both the distance between ions and beam waist radius of lasers. The Rabi frequency of laser can be regarded as constant in our scheme and cross-terms between lasers and ions are ignored. 
\begin{equation}
H_{2} =\sum_{n=1,2}\Omega_n\sin[\eta (a + a^\dagger ) + \phi_n]\sigma_{x,n} \cos\omega_{L,n} t,
\label{Eq3}
\end{equation}
where $ \Omega_n$, $\omega_L^n$ and $\phi_n$ are corresponding to the Rabi frequency, frequency and phase of $n$th 729~nm lasers, the Pauli operator of $n$th ion is $\sigma_{x,n} = |g\rangle_n\langle e|+|e\rangle_n\langle g|$, and $a$ ($a^\dagger$) is the annihilation (creation) operator of radial center-of-mass phonon mode of two ions. The Lamb-Dicke parameter of lasers is given as $\eta = k x_0\cos\alpha $, where the $k$ and $\alpha$ denote the wave vector and incident angle of lasers, and the zero-point fluctuation of the radial mode is $x_0=\sqrt{\hbar/4m\omega_r}$ with the Planck constant $\hbar$, the mass $m$ of a single ion and vibrational frequency $\omega_r$ of radial mode. Transferring the system into the rotating picture of $\omega_e a^{\dagger}a+\omega_{g,n}\sigma_{z,n}/2$, where $\omega_{g,n}$ is the energy difference between the ground state $|g\rangle_n$ and metastable state $|e\rangle_n$, and $\omega_e$ denotes the frequency of external alternating electric field, the Hamiltonian is then obtained as $H_2=\sum_{n=1,2}H_{2,n}$ with
\begin{equation}\label{Eq4}
H_{2,n}=\Omega_n\sin[\eta (ae^{-i\omega_et} + a^\dagger e^{i\omega_et}) + \phi_n](\sigma_{-,n} e^{i\delta_n t} + h.c.),
\end{equation}
where the detuning of laser is $\delta_n = \omega_{L,n} - \omega_{g,n}$ and the lower (upper) operator $\sigma_{-,n}=|g\rangle_n\langle e|$ ($\sigma_{+,n}=|e\rangle_n\langle g|$).

To obtain the necessary linear gain on the radial phonon mode, we select the phase of first laser as $\phi_1=0$ and then make the Lamb-Dicke approximation under the condition $\eta\sqrt{\langle n\rangle}\ll 1$ with $\langle n\rangle$ denoting the average phonon number of radial mode. Keeping to the third-order term of $\eta$, the Hamiltonian is obtained as $H_{2,1}=H_{2,1}^{(1)}+H_{2,1}^{(2)}$ with
\begin{equation}
\begin{aligned}
H_{2,1}^{(1)}&=  \lambda_1(ae^{-i\omega_et} + a^\dagger e^{i\omega_et}) (\sigma_{-,1} e^{i\delta_1 t} + h.c.), \\ \notag
H_{2,1}^{(2)}&= -\frac{\lambda_1\eta^2}{6}(ae^{-i\omega_et} + a^\dagger e^{i\omega_et})^3 (\sigma_{-,1} e^{i\delta_1 t} + h.c.), \notag
\end{aligned}
\end{equation}
where the linear coupling strength $\lambda_1=\eta\Omega_1$. Selecting $\delta_1 = \omega_e$ and making the rotating-wave approximation under the condition $\omega_e\gg \lambda_1$, they reduce to
\begin{equation}
\begin{aligned}
H_{2,1}^{(1)}&= \lambda_1(\sigma_{-,1} a+  \sigma_{+,1}a^\dagger)-\frac{\lambda_1^2}{2\omega_e}(a^{\dagger}a\sigma_{z,1}+\sigma_{+,1}\sigma_{-,1}), \\ \notag
H_{2,1}^{(2)}&=-\frac{\lambda_1\eta^2}{2}[\sigma_{-,1}a(a^{\dagger}a) +  \sigma_{+,1}a^\dagger (a^{\dagger}a+1)]. \notag
\end{aligned}
\end{equation}
The second term in $H_{2,1}^{(1)}$, obtained by the rotating-wave terms of first-order term, is a phonon number dependent Stark shift and % is intractable to compensate it by controlling the laser, fortunately, it 
can be dropped due to $\lambda_1\sqrt{\langle n\rangle}\ll\omega_e$. Besides, we can rewrite $H_{2,1}^{(2)}$ as $H_{2,1}^{(2)}=-\eta^2 \langle n\rangle H_{2,1}/2$, which induces a phonon number dependent decrease of coupling strength. % (this effect is also intractable to compensate it by controlling the laser). 
When $\eta^2 \langle n\rangle\ll 1$, we can further ignore the second term and obtain the Hamiltonian as
\begin{equation}\label{Eq5}
H_{2,1}= \lambda_1(a\sigma_{-,1} + a^\dagger \sigma_{+,1}).
\end{equation}
Further adiabatically eliminating the state $|e\rangle_1$ under the condition $\Gamma_1\gg \lambda_1$, the effective dynamics process can be obtained as (see Appendix.~\ref{PA}) 
\begin{equation}
L_{1,\text{eff}}= i\frac{2\lambda_1}{\sqrt{\Gamma_1}}a^{\dagger} |g\rangle_1\langle g|, \quad g = \frac{4\lambda_1^2}{\Gamma_1},
\label{Eq6}
\end{equation}
with the restriction condition $ \Gamma_1\gg \lambda_1\gg g$. Then, tracing over the ground state $|g\rangle_1$, the effective linear gain on the phonon mode is obtained with the gain rate $g$.

\begin{figure*}[htbp]
    \centering
    \includegraphics[width=17.0cm,height=5.5cm]{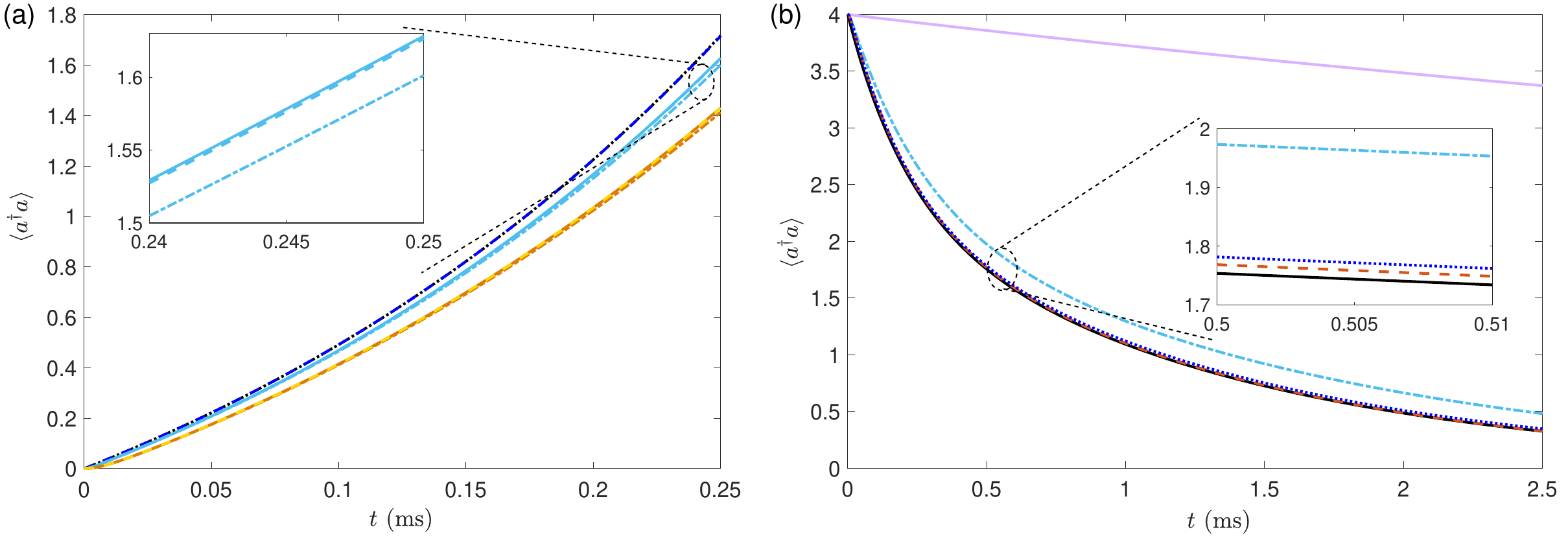}
    \caption{(a)The evolution of average phonon number in phonon mode for the given linear gain rate $g=4$~kHz, where the blue solid curve denotes analytical result, green dash curve is the result of effective dynamics under the master equation of Eq.~(\ref{Eq12}), the solid curves are the results under the master equation of Eq.~(\ref{Eq9}) with Hamiltonian given by Eq.~(\ref{Eq5}), the dot-dash curves are the results under the master equation of Eq.~(\ref{Eq9}) with Hamiltonian given by Eq.~(\ref{Eq10}) and the dash curves are obtained with the compensation of third-order term $H_{2,1}^{(2)}$, respectively. The initial state of phonon mode and spin state are corresponding to the Fock state $|0\rangle$ and ground state $|g\rangle$, the parameters are selected as $\Gamma=1.09$~MHz, $\lambda_1=33$~kHz and $\Omega_1=482$~kHz for the cyan curves, and they are selected as $\Gamma=272$~kHz, $\lambda_1=16.5$~kHz and $\Omega_1=241$~kHz for the brown curves. The red curve in the inset denote the results by consider the second term $H_{2,1}^{(2)}$ of $H_{2,1}$. (b) The evolution of average phonon number in phonon mode for the given nonlinear damping rate $\kappa=0.4$~kHz, where the black solid curve denotes the result of effective dynamics under the master equation of Eq.~(\ref{Eq12}), and other curves are obtained by the master equation of Eq.~(\ref{Eq9}), where the Hamiltonian is given by Eq.~(\ref{Eq7}) with $\lambda_2=\eta^2\Omega_2/2$ for the red dash curve, the Hamiltonian given by Eq.~(\ref{Eq11}) without Stark shift compensation and coupling strength amendment for the purple curve, with only Stark shift compensation for the cyan the dot-dash curve, and with Stark shift compensation and coupling strength amendment for the green dots curve. The initial state of phonon mode and spin state are corresponding to the Fock state $|4\rangle$ and ground state $|g\rangle$, the parameters are selected as $\Gamma_2=272.4$~kHz, $\lambda_2=5.22$~kHz, $\Omega_2=2.22$~MHz and $\bar{\delta}=787$~kHz without coupling strength amendment, and $\lambda_2=5.97$~kHz, $\Omega_2=2.54$~MHz and $\bar{\delta}=1.03$~MHz with the coupling strength amendment. Here the Lamb-Dicke parameter for two ions is $\eta=0.069$ and $\omega_r/2\pi=1$~MHz.}
    \label{Figs2}
\end{figure*}

To obtain the necessary nonlinear damping term on the phonon mode, we select the phase of second laser as $\phi_2=-\pi/2$ and make the Lamb-Dicke approximation under the condition $\eta\sqrt{\langle n\rangle}\ll 1$. Keeping to the first order term, we can obtain the Hamiltonian as
\begin{equation}
H_{2,2} \simeq -\Omega_2[1-\frac{\eta^2}{2} (ae^{-i\omega_et} + a^\dagger e^{i\omega_et})^2] (\sigma_{-,2} e^{i\delta_2 t} + h.c.). \notag
\end{equation}
Selecting the detuning of second laser as $\delta_2 =-2 \omega_e$ and making the rotating-wave approximation under the condition $\omega_e\gg \Omega_2$, it reduces to 
\begin{equation}
H_{2,2}= \lambda_2 (a^2\sigma_{+,2} + (a^\dagger)^2 \sigma_{-,2})-\Omega_2 (\sigma_{-,2} e^{-2i\omega_e t} + h.c.), \notag
\end{equation}
where the first term is necessary with the nonlinear coupling strength $\lambda_2=\eta^2\Omega_2/2$, while the second term is the undesired carrier transition. Since $ \Omega_2\gg \lambda_2$, the second term will induce a large Stark shift in the spin and a small amendment for the coupling strength. Keeping to the second-order approximation term of the carrier transition, the Hamiltonian is amended as,
\begin{equation}
H_{2,2}= \lambda_2(1-\frac{\Omega_2^2}{\omega_e^2}) [a^2\sigma_{+,2} + (a^\dagger)^2 \sigma_{-,2}]+\frac{\Omega_2^2}{2\omega_e} \sigma_{z,2}, \notag
\end{equation}
where the second term is the first-order approximation term of carrier transition, and the correction factor in the first term is the second-order approximation term of carrier transition. Since $\Omega_2^2/2\omega_e$ is in the same order with $\lambda_2$, we need to compensate the Stark shift. Thus, in the process of rotating picture for obtaining $H_{2,n}$ of Eq.~(\ref{Eq4}), we use $(\omega_{g,n}+\bar{\delta}_2)\sigma_{z,2}/2$ with the compensatory detuning $\bar{\delta}_2=\Omega_2^2/\omega_e$ to replace the original $\omega_{g,n}\sigma_{z,2}/2$, which amends the frequency of second laser as $\omega_{L,2}=\omega_{g,n}-2\omega_e+\bar{\delta}_2$, still keeping the detuning $\delta_2=-2\omega_e$. On the other hand, since the correction factor only influences the coupling strength, not producing extra prominent effect, we can drop this correction factor or rewrite it as $\lambda_2=\eta^2\Omega_2(\omega_e^2-\Omega_2^2)/2\omega_e^2$.  After compensating the Stark shift and dropping the correction factor, the Hamiltonian is reduced to 
\begin{equation}\label{Eq7}
H_{2,2}= \lambda_2 [a^2\sigma_{+,2} + (a^\dagger)^2 \sigma_{-,2}].
\end{equation}
Under the condition $\Gamma\gg \lambda_2$ and adiabatically eliminating the state $|e\rangle_2$, an effective dynamics process is obtained as (see Appendix.~\ref{PA}) 
\begin{equation}
L_{2,\text{eff}}=  i\frac{2\lambda_2}{\sqrt{\Gamma_2}}a^{2} |g\rangle_2\langle g|, \quad \kappa = \frac{4\lambda_2^2}{\Gamma_2},
\label{Eq8}
\end{equation}
with the restriction $ \Gamma_2\gg\lambda_2\gg \kappa$. After tracing over ground state $|g\rangle_2$, the effective nonlinear damping on the phonon mode is obtained with the rate $\kappa$.

To evaluate the validity of effective linear gain and nonlinear damping, we compare the master equation, consisting of both the spin state and phonon mode, with the efficient dynamics only consisting of the phonon mode. The master equation of compound system is described as following:
\begin{equation} \label{Eq9}
\dot{\rho}_{\text{sp},n}=-i[H_{\text{sp},n},\rho_{\text{sp},n}]+\frac{\Gamma_n}{2}\mathcal{D}[\sigma_{-,n}]\rho_{\text{sp},n},
\end{equation} 
where $\mathcal{D}[o]\rho=2o\rho o^{\dagger}-(o^{\dagger}o\rho+\rho o^{\dagger}o)$ and the subscript 's' and 'p' denote the spin state and phonon state, respectively. For the linear gain, the Hamiltonian is given as 
\begin{equation}\label{Eq10}
H_{\rm sp,1}=\omega_ra^{\dagger}a -\frac{\delta_1}{2}\sigma_{z,1}+\Omega_1\sigma_{x,1}\sin\eta (a + a^\dagger ),
\end{equation}
with the detuning $\delta_1=\omega_e$. For the nonlinear damping, the Hamiltonian is given as 
\begin{equation} \label{Eq11}
H_{\rm sp,2}=\omega_ra^{\dagger}a -\frac{\delta_2}{2}\sigma_{z,2}-\Omega_2\sigma_{x,2}\cos\eta (a + a^\dagger ),
\end{equation}
with the detuning $\delta_2=-2\omega_e+\bar{\delta}_2$. There are no rotating-wave approximation and Lamb-Dicke approximation, thus, reflecting the practical dynamics of system. For the efficient dynamics, it is described as 
\begin{equation}\label{Eq12}
\dot{\rho}_{\text{p},n}=-i[H_{\text{p}},\rho_{\text{p},n}]+\frac{1}{2}\mathcal{D}[L_{n,\rm eff}]\rho_{\text{p},n},
\end{equation}
where the Hamiltonian $H_{\rm p}=-\Delta a^{\dagger}a$ with the detuning as $\Delta=-(\omega_r-\omega_e)$, the linear gain operator $L_{1,\rm eff}=\sqrt{g}a^{\dagger}$ and nonlinear damping operator $L_{2,\rm eff}=\sqrt{\kappa}a^{2}$.

In Fig.~\ref{Figs2}, we compare the results under the effective dynamics and master equation dynamics with the original Hamiltonian, respectively. For the linear gain of phonon, Fig.~\ref{Figs2}(a) shows that the results of master equation can conform the effective dynamics with a high fidelity. The deviation between effective dynamics and master equation dynamics with Hamiltonian given by Eq.~(\ref{Eq5}) derive from the adiabatic elimination that a large decay rate $\Gamma$ from the state $|e\rangle$ to $|g\rangle$ can make the result of master equation more close to the effective dynamics since the ratio $\Gamma/\Omega_2$ increase as the increase of $\Omega_2$ for a given $g$, which makes the effective dynamics more valid. The other deviation comes from the Lamb-Dicke approximation that the increase of mean phonon number will break the Lamb-Dicke condition, i.e., $\eta\sqrt{\langle n\rangle}\ll 1$, which can be more well fitted after considering the high-order term (see the inset of Fig.~\ref{Figs2}(a)). On the other hand, we can also find that the effective dynamics is well coincided with the analytical result, given by $\langle a^{\dagger}a\rangle=(\langle a^{\dagger}a\rangle_0+1)e^{gt}-1$ with $\langle a^{\dagger}a\rangle_0$ being the initial average phonon number of phonon mode. For the nonlinear damping of phonon, Fig.~\ref{Figs2}(b) also shows that the results of master equation can conform the effective dynamics with a high fidelity. Different from the linear gain situation, the deviation between them can become very small and a higher fidelity can be reached by compensating the Stark shift and amending the coupling strength. 

In the third step, we add the Hamiltonian of external electric field to the system, and the Hamiltonian of the phonon mode is then obtained as
\begin{align}
H_a = \omega_r a^\dagger a -2q E x_0 (a +a^{\dagger})\cos(\omega_e t+\varphi), \notag
\end{align}
where $q$ represents the elementary charge of ions,  $\varphi$ and $E$ are corresponding to the phase and amplitude of the electric field. In the interaction picture of $\omega_e a^\dagger a$, the Hamiltonian can be rewritten as
\begin{align} \label{Eq13}
    H_a=-\Delta a^\dagger a +  \varepsilon a^\dagger+ \varepsilon^* a,
\end{align}
where the high-frequency terms ($e^{\pm 2\omega_e t}$) only induce a small constant energy translation $-|\varepsilon|^2/2\omega_e $, and we thus ignore the high-frequency terms. The detuning of driving field is $\Delta=\omega_e-\omega_r$ and we rewrite the drive strength as $\varepsilon = -q E  x_0 \mathrm{e}^{-i\varphi}$. Adding the linear gain and nonlinear damping to the phonon mode, the effective master equation can be obtained as
\begin{align}\label{Eq14}
    \dot{\rho} = -i[H_a,\rho] +  \frac{g}{2} \mathcal{D}[a^\dagger]\rho +\frac{\kappa}{2}   \mathcal{D}[a^2]\rho,
\end{align}
where the linear gain rate is $g = 4\eta^2 \Omega_1^2\gamma/\Omega_{e,1}^2$ and the nonlinear damping rate is $\kappa =\eta^4 \Omega_2^2\gamma/\Omega_{e,2}^2$.

\section{experimental feasibility and numerical simulation}\label{section3}

For the trapped $^{40}$Ca$^{+}$ ion system, the life time of excited state $4^2P_{3/2}$ is 6.9~ns, which gives a decay rate $\gamma / 2\pi = 22.4~\text{MHz}$~\cite{PhysRevLett.128.050603}. The trap frequency is accessible in an interval $\omega_r/2\pi\in [0.1, 10]$~MHz~\cite{RMP-93-025001}. The 854~nm laser can drive a dipole transition between the metastable state and the excited state which produces a Rabi frequency of several MHz, even tens of MHz. The 729~nm laser drives a quadrupole transition between the metastable state and ground state, which gives a maximal Rabi frequency generally around several hundred kHz and rarely reaches the MHz level~\cite{RMP-93-025001}. Particularly, the Lamb-Dicke parameter of a single $^{40}$Ca$^{+}$ ion is $\eta \approx 0.1$ for a trap frequency of $\omega_r/2\pi = 1~\text{MHz}$ and it decreases to $\eta=0.069$ for two ions. % As the number ($N$) of ions increases, $\eta$ of the collective vibrational modes diminishes proportionally to $1/\sqrt{N}$. Consequently, for the radial center-of-mass mode of two ions, the Lamb-Dicke parameter $\eta \in [0.022, 0.22]$ when the trap frequency is tuned within $\omega_t/2\pi \in [0.1, 10]~\text{MHz}$, which can be selected to optimize ion coupling. 

In the process of obtaining the linear gain and nonlinear damping processes on the vibrational mode, we have successively made different adiabatic approximations and neglect higher-order term, which restrict the involved parameters to satisfy the following relations
\begin{equation}\label{Eq15}
\begin{aligned}
&\gamma \gg \Omega_{e,n}\gg \Gamma_{n}, \Omega_{n}; \quad \omega_r \gg \Omega_{n}; \quad \eta \ll 1\\
&\Gamma_{n} \gg \eta \Omega_1, \eta^2 \Omega_2; \quad \eta \Omega_1 \gg g; \quad \eta^2 \Omega_2\gg \kappa,
\end{aligned}
\end{equation}
with $n=1,2$. These restricted conditions may limit the accessible parameters in a small range, and we will illustrate our scheme under the practically feasible parameters in the following discussion.

\begin{figure}[htbp]
    \centering
    \includegraphics[width=8.5cm,height=9.0cm]{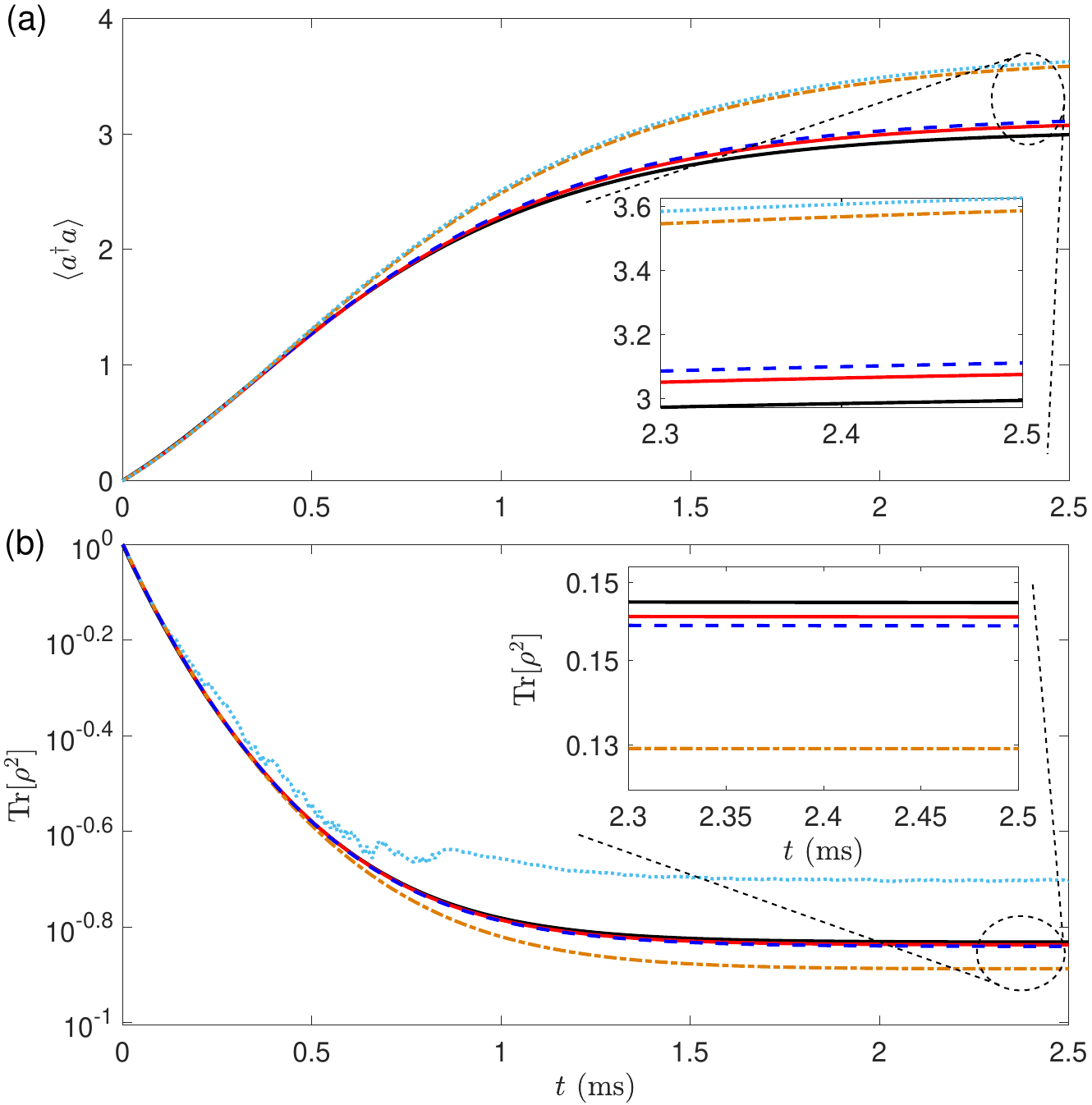}
    \caption{The evolution of average phonon number $\langle a^{\dagger}a\rangle$ (a) and the state purity $\text{Tr}(\rho^2)$ (b) in radial normal phonon mode of two ions, where the black solid and red solid curves are corresponding to the results under the master equation of Eq.~(\ref{Eq14}) and Eq.~(\ref{Eq17}), and cyan dots, brown dot-dash, green dash curves denote the results under the master equation of Eq.~(\ref{Eq16}) with only the compensation of Stark shift, with the compensation of third-order term $H_{2,1}^{(2)}$ and Stark shift, and with the compensation of third-order term $H_{2,1}^{(2)}$, Stark shift and coupling strength amendment, respectively. The initial state of phonon mode and spin state for the two ions are corresponding to the vacuum state $|0\rangle$ and ground state $|g\rangle_1|g\rangle_2$. The parameters are selected as $\Gamma_1=1.09$~MHz, $\lambda_1=23.3$~kHz, $\Omega_1=341$~kHz, $\Gamma_2=272$~kHz, $\lambda_2=5.22$~kHz, $\Omega_2=2.22$~MHz and $\bar{\delta}=787$~kHz without coupling strength amendment, and $\lambda_2=5.97$~kHz, $\Omega_2=2.54$~MHz and $\bar{\delta}=1.03$~MHz with the coupling strength amendment. Here the Lamb-Dicke parameter for two ions is $\eta=0.069$, the vibrational frequency $\omega_r/2\pi=1$~MHz, the linear gain $g=2$~kHz, the nonlinear damping $\kappa=0.4$~kHz, and the frequency of external electric field is selected as $\omega_e=\omega_r$ and the driving strength $\varepsilon=0$, respectively. }
    \label{figs3}
\end{figure}

To further evaluate the validity of effective dynamics to obtain the time crystal in the phonon mode, we use the preliminary master equation of compound system (consisting of spin and phonon mode) without any approximations to numerically calculate the evolution of phonon mode, described as 
\begin{equation} \label{Eq16}
\dot{\rho}_{t}=-i[H_{t},\rho_{t}]+\frac{\gamma}{2}\sum_{n=1,2}\mathcal{D}[\tilde{\sigma}_{-,n}] \rho_{t},
\end{equation} 
with the operator $\tilde{\sigma}_{-,n} = |e\rangle_n \langle p|$ and the total  Hamiltonian of compound system as
\begin{equation}
\begin{aligned}
H_t=& \omega_r a^\dagger a-\frac{\omega_e}{2}\sigma_{z,1}+\frac{2\omega_e-\bar{\delta}_2}{2}\sigma_{z,2}+\frac{\Omega_e}{2}\sum_{n=1,2}\tilde{\sigma}_{x,n} \\
&+(\varepsilon^*ae^{i\omega_e t}+\varepsilon a^{\dagger}e^{-i\omega_e t})+\Omega_1\sigma_{x,1}\sin\eta (a + a^\dagger ) \\
&-\Omega_2 \sigma_{x,2}\cos\eta (a + a^\dagger), \notag
\end{aligned}  
\end{equation}
where we have ignored the small constant energy translation induced by the high-frequency components of driving terms, and the state of phonon mode is obtained as $\rho=\text{Tr}_s(\rho_t)$ by tracing over the spin state of compound system.  

Apart from the preliminary master equation in Eq.~(\ref{Eq16}), as shown in Fig.~\ref{Figs2}, we have also shown that the simplified effective dynamics of master equation under the rotating-wave approximation and Lamb-Dicke approximation can also effectively describe the evolution of compound system with a high fidelity if the lower phonon condition $\eta\sqrt{\langle n\rangle}\ll 1$ is satisfied, and the simplified effective master equation can be obtained as 
\begin{equation} \label{Eq17}
\dot{\rho}_{e}=-i[H_{e},\rho_{e}]+\frac{\Gamma}{2}\sum_{n=1,2}\mathcal{D}[\sigma_{-,n}]\rho_{e},
\end{equation} 
where $\Gamma_n=\Omega_{e,n}^2/\gamma$, $\sigma_{-,n} = |g\rangle_n \langle e|$ and the effective Hamiltonian 
\begin{equation}\label{Eq18}
\begin{aligned}
H_e=& -\Delta a^\dagger a+(\varepsilon a^\dagger+ \varepsilon^* a)+\lambda_1(a\sigma_{-,1}+  a^\dagger \sigma_{+,1}) \\
&+\lambda_2 [a^2\sigma_{+,2} + (a^\dagger)^2 \sigma_{-,2}].
\end{aligned}  
\end{equation}

In the Fig.~\ref{figs3}, it shows that the effective dynamics of phonon mode can be well approximated by the master equation of Eq.~(\ref{Eq17}) under the restricted condition, meanwhile, the dynamics of preliminary master equation in Eq.~(\ref{Eq16})  can also be well fitted with the effective dynamics after making the corresponding coupling strength amendment and compensation of Stark shift and third-order approximation. We can also see that the third-order term of Lamb-Dick approximation to obtain the linear gain and the second-order amendment of coupling strength to obtain the nonlinear damping have an increasing influence on the dynamics as the increase of phonon number, thus, it is necessary to consider these high-order effect when the average phonon number become large. In the following, to simplify the process of demonstrating the time crystal of vibrational mode, we choose the frequency of radial center-of-mass mode as $\omega_r/2\pi=1$~MHz, giving the Lamb-Dicke parameter of two ions $\eta = 0.0685$, and setting the linear gain $g=2$~kHz. Then, we adjust the detuning $\Delta$ and strength $\varepsilon$ of external electric field, and the nonlinear damping $\kappa$ (i.e., the corresponding parameters $\lambda_2$, $\Omega_2$, $\delta_2$ and $\Gamma_2$) to demonstrate the time crystal of phonon mode.

\begin{figure*}[htbp]
    \centering
    \includegraphics[width=17.5cm,height=10.0cm]{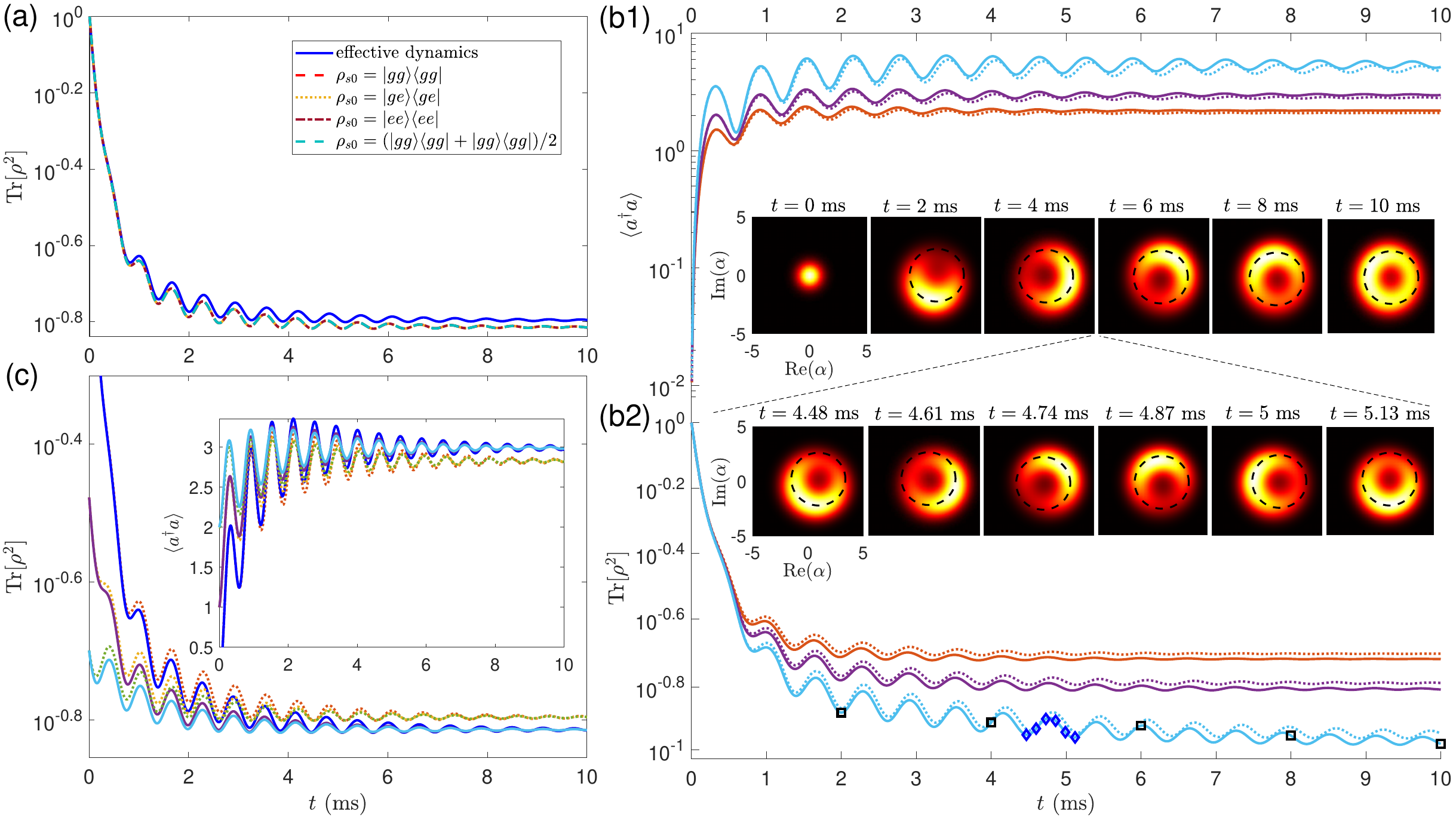}
    \caption{(a) The evolution of state purity $\text{Tr}(\rho^2)$ under different initial states in the spin states of two ions, where the initial state of phonon mode is in the vacuum state $|0\rangle$, the blue solid curve denotes the effective dynamics under the master equation of Eq.~(\ref{Eq14}), and other curves are obtained by the master equation of Eq.~(\ref{Eq17}). (b1) and (b2) The evolution of average phonon number $\langle a^{\dagger}a\rangle$ and state purity $\text{Tr}(\rho^2)$ for the different nonlinear damping rates $\kappa=0.6, 0.4, 0.2$~kHz corresponding to the red, purple and cyan curves, where the dot and solid curves are obtained by the master equation of Eq.~(\ref{Eq14}) and Eq.~(\ref{Eq17}), respectively. The insets of (b1-b2) denote the quantum Husimi distribution of phonon states at different evolution time (shown by the square and diamond dots in the cyan curve), where the red dashed line represents the limit cycle. (c) The evolution of the state purity $\text{Tr}(\rho^2)$ for the different thermal state of phonon mode, where the curves from the upper to bottom are corresponding to the initial average phonon number $\langle a^{\dagger}a\rangle_{\rm ini}=0, 1, 2$, the dot and solid curves are obtained by the master equation of Eq.~(\ref{Eq14}) and Eq.~(\ref{Eq17}), respectively, and the inset denotes the corresponding evolution of average phonon number $\langle a^{\dagger}a\rangle$. Here, we obtain the nonlinear damping $\kappa=0.6, 0.4, 0.2$~kHz by adjusting the $\Omega_2$ and $\delta_2$ for the given $\Gamma_2=272$~kHz, the parameters are selected as $\Gamma_1=1.09$~MHz, $\Omega_1=341$~kHz and $\lambda_1=23.3$~kHz to obtain the linear gain $g=2$~kHz, the frequency of external electric field is selected as $\Delta=5g$ and the driving strength $\varepsilon=\sqrt{g(g^2+4\Delta^2)}/8\sqrt{\kappa}$, respectively, and the duration of evolution $\tau=20/g$. }
    \label{figs4}
\end{figure*}

\section{Phonon time crystal in the vibrational mode}\label{section4}
\subsection{Classical driven Van der Pol oscillator model}
To further understand the dynamics of the quantum system, we first consider the dynamics in the thermodynamic limit, where the Lindblad master equation can be further reduced to a driven classical Van der Pol oscillator model with the evolution of amplitude $\alpha = \left< a \right>$ described by the following differential equation (see Appendix~\ref{PB})
\begin{align}\label{Eq19}
    \dot{\alpha} = \left ( \frac{g}{2} + i\Delta -\kappa |\alpha|^2\right ) \alpha - i \varepsilon.
\end{align}
It exhibits both the Hopf bifurcation and a limit cycle phase~\cite{VDP1,VDP2,VDP3,VDP4}. In the limit of $\kappa \to 0$~\cite{thermodynamiclimit}, the Hopf bifurcation point can be characterized by the rescaled driving strength parameter $ \varepsilon \sqrt{\kappa} = \sqrt{g(g^2 + 4\Delta^2)}/4$. This classical model is well studied and we focus on investigating the nonequilibrium behaviour at the full quantum level, especially in the phonon mode of a quantum harmonic oscillator.

\subsection{Phonon time crystal of ions}
To study the phonon time crystal of ions, as shown in Fig.~\ref{figs4}, we numerically solve the time evolution of phonon number and purity in the vibrational mode. Firstly, since both the linear gain and nonlinear damping are obtained by the assist of spin state, we study the influence of initial spin states on the dynamics, as shown in Fig.~\ref{figs4}(a), which well demonstrates a spin-state independent dynamics and we set the initial state of spins as $|gg\rangle$ for simplicity. Then, in Fig.~\ref{figs4}(b1-b2), we demonstrate the time evolution of phonon mode under different nonlinear damping rate, where as the nonlinear damping rate becomes weaker, two notable changes occur: an increase in the amplitude of oscillations and a lengthening of the relaxation time. These observations highlight the dynamic behaviour of the system as it passes through different regimes. The oscillatory nature of the evolution provides the strong evidence for the dissipative gap closing in the thermodynamic limit, thereby confirming that the vibrational mode behaves as a time crystal. Besides, we further analyse the time-dependent evolution of the state purity that at the beginning of the evolution, the purity rapidly drops with the similar rate due to the system initially in the nonequilibrium state, and then the system enters a metastable state after this initial decrease, where the purity temporarily stabilizes before the following oscillations. Over time, the oscillations gradually dampen, and the system approaches a steady-state value of purity.

To further depict the dynamics of time crystal, we study the evolution process of Husimi $Q$ function defined as $    Q(\alpha) =\langle \alpha | \rho | \alpha \rangle/\pi$, where $| \alpha  \rangle $ is a coherent state in phase space with the complex number $\alpha = q + ip$ corresponding to a specific point in the phase space~\cite{Husimi1940264}. The Husimi $ Q$ function provides a quasi-classical representation of the quantum state, allowing us to intuitively understand how the quantum state is distributed in the classical phase space, and also provides a quasiprobability distribution with the maximal value partially reflecting the phase fluctuation of the quantum oscillation. Similar to the evolution of state purity, starting from the initial vacuum state, there is a rapid dephasing process at beginning, followed by a slowdown in the rate of dephasing, and then continuously oscillating~\cite{VDP1}. Eventually, the oscillations gradually fade and the quantum limit cycle becomes blurred. In the inset of Fig.~\ref{figs4}(b1-b2), we illustrate the evolution of the $Q$ function: Initially, the $Q$ function appears as a Gaussian wave packet, reflecting the coherence of the initial state and aligned with the initial state of our setup; As the evolution progresses, the $Q$ function gradually disperses, where the wave packet rotates along the limit cycle in phase space and is increasingly elongated; In a stable time crystal phase, it shows the stable quantum oscillation.

%control error
\begin{figure}[htbp]
    \centering
    \includegraphics[width=8.4cm,height=5.4cm]{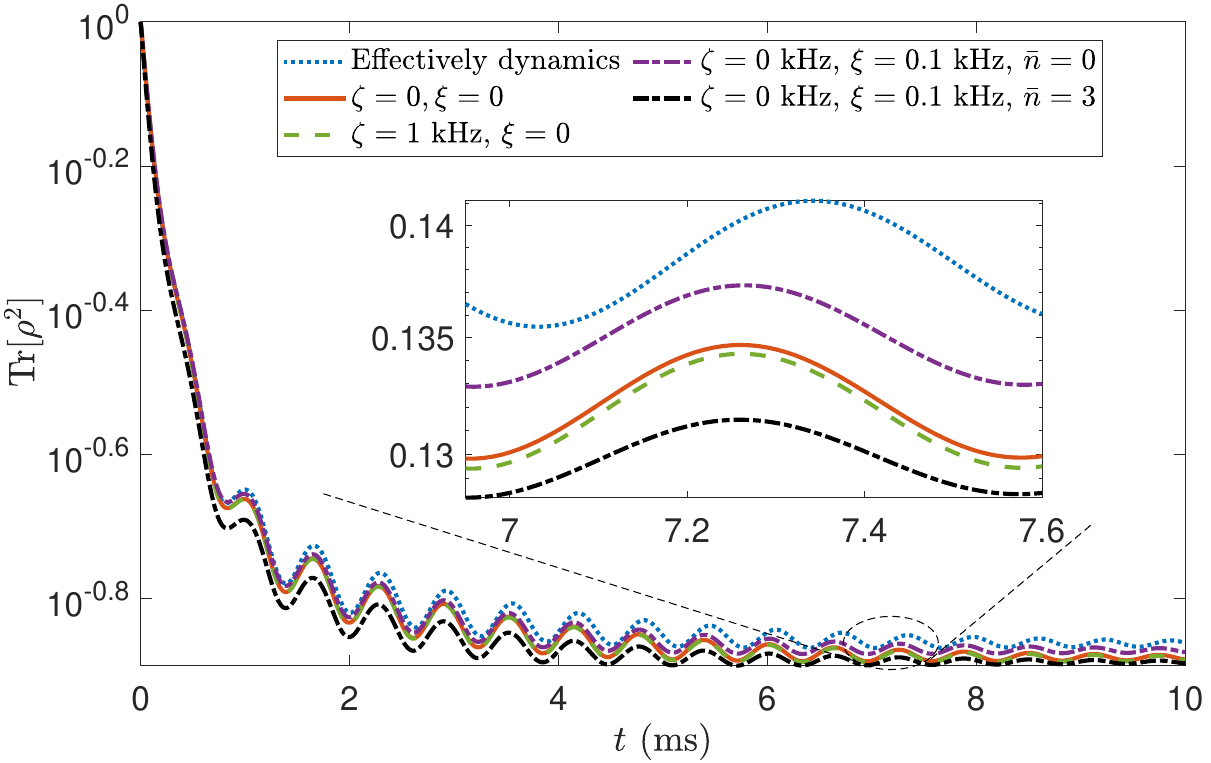}
    \caption{The evolution of state purity $\text{Tr}(\rho^2)$ under the spin dephasing and thermalization of phonon mode, where the dephasing rates of spin state are set $\zeta_1=\zeta_2=\zeta$ and the heating rates of phonon mode are given as $ \xi$.  Here, the given Rabi frequencies are $\Omega_1=341$~kHz and  $\Omega_2=2.22$~MHz, the nonlinear damping $\kappa= 0.4$~kHz, the linear gain rate $g=2$~kHz, the frequency of external electric field is selected as $\Delta=5g$ and the driving strength $\varepsilon=\sqrt{g(g^2+4\Delta^2)}/8\sqrt{\kappa}$, respectively, and the duration of evolution $\tau=20/g$. }
    \label{figs6}
\end{figure}

\subsection{Robustness to the thermalization of phonon mode and  dephasing of spin states}\label{section5}
In the practical experiments, the system cannot always be prepared in the vacuum state as described in Sec.~\ref{section4} or an ideal coherent state, but the state of the system more generally evolves into a thermal state with an effective temperature. Consequently, the impact of thermal effects on the evolution of the time crystal will be considered. To discuss the thermal effect, we assume that we initially prepare the vibrational mode in a thermal state $\rho_{\rm th}=\exp(-\beta H_{r0})/\text{Tr}[\exp(-\beta H_{r})]$ with the inverse temperature $\beta=1/k_B T$ ($k_B$ is the Boltzmann constant and $T$ is the effective temperature of vibrational state) and Hamiltonian of vibrational mode $H_{r}=\hbar\omega_ra^{\dagger}a$~\cite{Strocchi2008,Guryanova2016}, which gives an initial average phonon number $\langle a^{\dagger}a\rangle_{\rm ini}=1/(e^{\beta \hbar\omega_r}-1)$ and a probability distribution on the Fock state $|n\rangle$ as $p_n=\langle a^{\dagger}a\rangle_{\rm ini}^n/(\langle a^{\dagger}a\rangle_{\rm ini}+1)^{n+1}$. As shown in Fig.~\ref{figs4}(c), our model demonstrates remarkable robustness against the different thermal states for the stable quantum oscillation, i.e., the choice of the thermal state does not affect the oscillation period of the time crystal but only leads to changes in the amplitude. As a result, we are able to directly use the thermal state as the initial condition. 

%control error
\begin{figure*}[htbp]
    \centering
    \includegraphics[width=15.4cm,height=4.8cm]{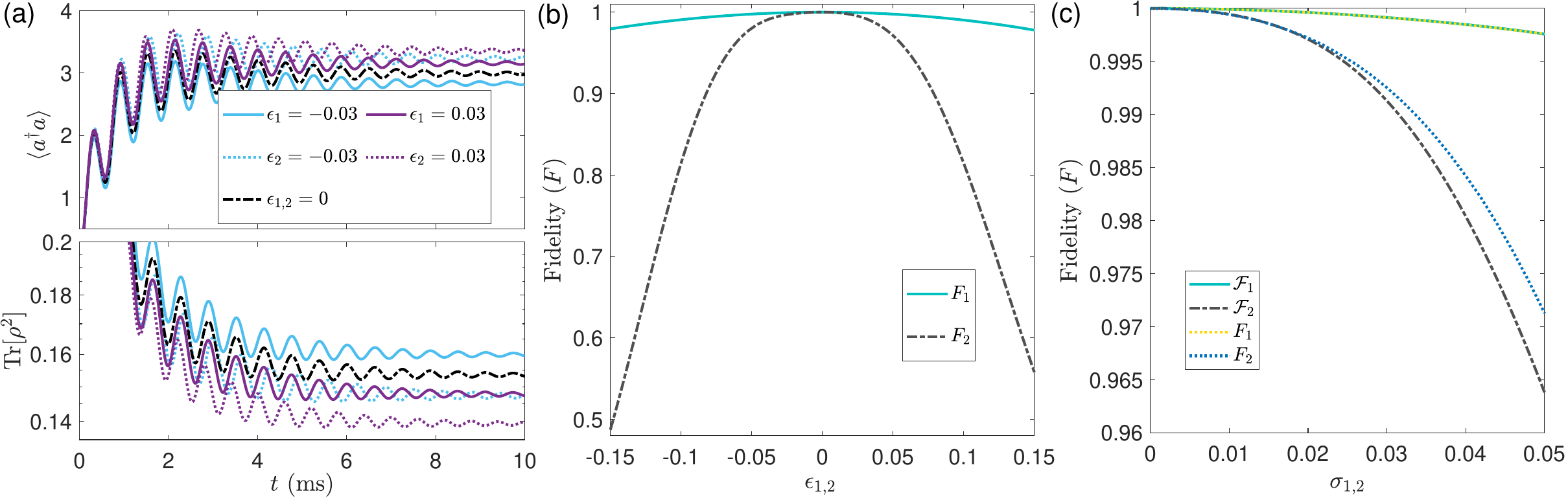}
    \caption{(a) The evolution of average phonon number $\langle a^{\dagger}a\rangle$ and state purity $\text{Tr}(\rho^2)$ under different control errors, where the numerical results are obtained by the master equation of Eq.~(\ref{Eq16}) with the compensation of third-order term $H_{2,1}^{(2)}$, Stark shift and coupling strength amendment. (b) The fidelity $F$ of evolution process between the dynamics with and without the given control deviations. (c) The fidelity $\mathcal{F}$ of evolution process for a given variance of control errors under the Gaussian noise. Here, the other parameters are selected same as that in Fig.~\ref{figs6}. }
    \label{figs5}
\end{figure*}

Then, we further consider the thermalization process of phonon mode, which generally induces the heating effect on the phonon mode, described by the additional dissipation term as 
\begin{equation} \label{Eq20}
L_T (\rho)=\frac{\xi}{2} [(\bar{n}+1)\mathcal{D}(a)\rho+ \bar{n} \mathcal{D}(a^{\dagger})\rho],
\end{equation}
where $\xi_j$ is the dissipative rate and $\bar{n}$ is the average phonon number of thermal equilibrium state with environment, given as $\bar{n}=[\exp(\hbar \omega_x/k_B T)-1]^{-1}$ with the temperature $T$ of environment and Boltzmann constant $k_B$. In the ion system, the temperature of environment always regarded as zero, i.e., $T=0$, which gives $\bar{n}=0$. In the practical experiment, the accessible heating rate is always smaller than $100$~Hz. Under these conditions, the thermalization process of phonon mode always has a negligible influence on the phonon time crystal. As shown in the Fig.~\ref{figs6}, it shows the robustness to the thermalization of phonon that the thermalization process doesnot change the characteristics of phonon time crystal, i.e., the stable periodic oscillation.

Finally, we consider the spin dephasing influence on the phonon time crystal, which adds the additional dephasing terms
\begin{equation}\label{Eq21}
L_z (\rho)=\sum_{j=1,2}\frac{\zeta_j}{2} (\sigma_{z,j} \rho \sigma_{z,j}-\rho),
\end{equation}
where $ \sigma_{z,j}$ and $\zeta_j$ denote the Pauli operator and dephasing rate of $j$th ion, respectively. In the $ ^{40}$Ca$^+$ ion system, the lifetime of metastable states is about $1$~s, giving the decay rate about $1$~Hz. The dephasing process of optical qubit is induced by the noise of magnetic field, generally, a dephasing time longer than $1$~ms can be easily obtained in the experiment, thus, giving the dephasing rate smaller than $1$~kHz. As shown in the Fig.~\ref{figs6}, the small spin dephasing also only produces a negligible deviation and doesnot change the characteristics of phonon time crystal. Thus, this scheme is also robustness to the spin dephasing process. 

\subsection{Robustness to the control errors of lasers}\label{section6}
On the other hand, there are always control errors for the Rabi frequencies and detunings of control lasers in the actual experiments. In our scheme, the Rabi frequencies of control lasers, especially two standing-wave lasers, determining the linear gain and nonliear damping, are more important than other parameters. To assess the robustness to the control errors, we first define the ratio of control deviation as $\epsilon_k=(\tilde{\Omega}_k-\Omega_k)/\Omega_k$, where $k=1,2$, $\Omega_k$ and $\tilde{\Omega}_k$ denote the given Rabi frequency and actual Rabi frequency of laser. In Fig.~\ref{figs5}(a), it shows that $\epsilon_1<0$ and $\epsilon_1>0$ shows the different departure directions from the situation without deviation, i.e., distributed on both sides of no deviation dynamics, while $\epsilon_2$ always induces the departure along the same direction from the situation without deviation. Besides, it also shows that the departure induced by $\epsilon_2$ is larger than that induced by $\epsilon_1$, which comes from that the control deviation of $\epsilon_k$ not only affect the effective nonliear damping rate, but also produces a prominent deviation on the detuning induced by the carry term. This results can also be seen in Fig.~\ref{figs5}(b-c), where we have used the mean fidelity $F(t)=\tau^{-1}\int_0^{\tau}F(t)dt$, defined in Sec.~\ref{section2}, to evaluate the fidelity of evolution process. Thus, the phonon time crystal  has stronger robustness to the linear gain error than the nonliear damping error. 

On the other hand, the control errors of lasers are always stochastic, not a fixed deviation, and we further assume the control errors induced by the Gaussian noise with the errors distribution as $p(\epsilon_{k})=e^{-\epsilon_{k}^2}/\sqrt{2\pi}\sigma_{k}$, where $k=1,2$ and $\sigma_{k}$ denotes the variance of control errors under the Gaussian noise. The average fidelity under the stochastic errors is given as $\mathcal{F}_k=\int_{-\infty}^{\infty}F(\epsilon_{k})p(\epsilon_{k})d\epsilon_{k}$. Since the larger deviation induces a smaller fidelity (see Fig.~\ref{figs5}(b)), the average fidelity $\mathcal{F}_k$, as shown in Fig.~\ref{figs5}(c), is always smaller than the fidelity under the same given deviation. Besides, it also shows that the decrease of average fidelity $\mathcal{F}_k$ is small within a small Gaussian noise and the phonon time crystal behaviour is still robustness to the control errors. Actually, the control errors can generally be limited in $2\%$ even more excellent in the trapped ions system, thus, our scheme is feasible and it is possible to experimentally observe the phonon time crystal of ions.

\section{Conclusion}\label{section6}

In summary, we have constructed a continuous-time crystal in a linear ion trap by employing two ions. Compared to the existing methods for implementing discrete-time crystals in ion traps, our scheme offers the enhanced stability for the time crystal behaviour. Based on the practical parameters and detailed numerical simulations, we have verified the experimental feasibility of this scheme and observed the time-dependent evolution results for the system's phonon number, purity, and $Q$-function under the accessible experimental conditions and parameters, confirming the existence of the time crystal. Furthermore, we demonstrate that our continuous-time crystal model exhibits remarkable robustness against the initial thermal states, thermalization and control errors of lasers. Our findings provide a feasible approach for constructing time crystals and offer valuable insights for research in many-body physics and non-equilibrium thermodynamics.

\section*{Acknowledgments}
This work is supported by the National Key Research and Development Program of China under Grant Nos. 2022YFA1404500 and 2021YFA1400902, by Cross-disciplinary Innovative Research Group Project of Henan Province under Grant No. 232300421004, National Natural Science Foundation of China under Grant Nos. 1232410, U21A20434, 12074346, 12274376, 12374466, 12074232, 12125406, by Natural Science Foundation of Henan Province under Grant Nos. 232300421075, 242300421212, by Major science and technology project of Henan Province under Grant No. 221100210400. \\

\appendix

\section{Adiabatic Elimination of Spin States} \label{PA}
To produce the effective dynamics, we first rephrase the effective operator formalism in Ref.~\cite{EffH}. Assuming the density matrix $\rho$ of the system with the Hamiltonian $H$ can be described by the Lindblad master equation as
\begin{equation}
    \dot{\rho} = -i\left[H,\rho \right] + \frac{1}{2}\sum_{n=1}^N [2L_n \rho L_n^\dagger -(L_n^\dagger L_n \rho + \rho L_n^\dagger L_n)], \notag
\end{equation}
where the Hamiltonian $H=H_g+H_e+V_{+}+V_{-}$ consists of the ground-state subspace Hamiltonian $H_g$, the excited-state subspace Hamiltonian $H_e$ and the coupling terms $V_+\ (V_-)$ between the excited-state subspace and ground-state subspace of system, and $L_n$ denotes the $n$th dissipative process from the excited-state subspace to ground-state subspace of system. Then, the effective dynamics of system after tracing out the excited-state subspace can be described as
\begin{equation}
    \dot{\rho} = -i\left[H_{\text{eff}},\rho \right] + \frac{1}{2}\sum_{n=1}^N \mathcal{D}[L_{n,\text{eff}}] \rho,
\end{equation}
where the reduced effective Hamiltonian and Lindblad operators are given as
\begin{align}
    H_{\text{eff}} &= -\frac{1}{2} V_{-} \left[H_{\text{NH}}^{-1} + (H_{\text{NH}}^{-1})^\dagger\right]V_{+} + H_g,\\
    L_{n,\text{eff}} &= L_n H_{\text{NH}}^{-1} V_{+},
\end{align}
with the non-Hermitian Hamiltonian$H_{\text{NH}}$ given as
\begin{equation}\label{HNH}
    H_{\text{NH}} = H_e - \frac{i}{2}\sum_k L_k^\dagger L_k.
\end{equation}
During this adiabatic elimination process, the dynamics of excited-state subspace should be much faster than the dynamics of  ground-state subspace and the coupling dynamics between them, meanwhile, the coupling dynamics between them should be much faster than the dynamics of ground-state subspace.

In the first step, the dynamics of system can be described by the Lindblad master equation as
\begin{equation}
    \dot{\rho} = -i\left[H_1,\rho \right] + \frac{1}{2}\sum_{n=1}^2 [2L_n \rho L_n^\dagger -(L_n^\dagger L_n \rho + \rho L_n^\dagger L_n)], \notag
\end{equation}
with the dissipative operator of $n$th ion as $L_n = \sqrt{\gamma}|g\rangle_n\langle p|$. Our aim is eliminating the excited state $|p\rangle$ to reduce the system's dynamics to an effective dynamics. In this situation, we first change the system in the rotating picture of $\omega_{e,n}\tilde{\sigma}_{z,n}/2$ and obtain the Hamiltonian as $H_{1} = \Omega_{e,n}\sum_{n=1,2} \left( \tilde{\sigma}_{-,n} + h.c. \right)/2$. Then, the Hamiltonian is redefined as $H_1=  H_g+H_e +V_-+V_+$ where the ground-state subspace Hamiltonian $ H_g= 0$, the excited-state subspace Hamiltonian $H_e=0$, the coupling terms between them $ V_-=\sum_{n = 1}^2  \Omega _{e,n}\tilde{\sigma}_{-,n}/2$ and $ V_+=\sum_{n = 1}^2  \Omega _{e,n}\tilde{\sigma}_{+,n}/2$, and the Lindblad operators $L_n = \sqrt{\gamma} |0 \rangle_n \langle 2 |$. Using Eq.~(\ref{HNH}), we can obtain
\begin{equation}
    H_{\text{NH}} = -\frac{i \gamma}{2} \sum_{n=1}^2\left| p \rangle_n \langle p \right|, \quad    H_{\text{NH}}^{-1} = \frac{2i}{\gamma}  \sum_{n=1}^2\left| p \rangle_n \langle p \right|,
\end{equation}
which result into
\begin{equation}
    H_{\text{eff}}= 0, \quad    L_{n,\text{eff}}= i\frac{\Omega_{e,n}}{\sqrt{\gamma}}\sigma_{-,n},
\end{equation}
where the lower operator of $n$th ion is given as $\sigma_{-,n}=|g\rangle_n\langle e|$. Thus, an effective decay rate is given as $\Gamma_n=\Omega_{e,n}^2/\gamma$ from the metastable state $|e\rangle_n$ to ground state $|g\rangle_n$ and the Stark shift induced by the laser is also eliminated. Moreover, the constraint conditions for adiabatic elimination mean that $\gamma\gg\Omega_{e,n}\gg\Gamma_n$.

In the second step, the Hamiltonian to produce the effective linear gain is given by Eq.~(\ref{Eq5}), which can also be written as $H_{2,1}= H_a+ H_g+H_e +V_-+V_+$ with the ground-state subspace Hamiltonian $ H_g=0$, the excited-state subspace Hamiltonian $H_e=0$, the coupling terms between them $ V_-= \lambda_1 a \sigma_{-,1}$ and $ V_+= \lambda_1 a^{\dagger} \sigma_{+,1}$, and the Lindblad operators $L = \sqrt{\Gamma} |0 \rangle_1 \langle e|$. Using Eq.~(\ref{HNH}), we can obtain
\begin{equation}
    H_{\text{NH}} = -\frac{i \Gamma}{2} \left| e \rangle_1 \langle e \right|, \quad    H_{\text{NH}}^{-1} = \frac{2i}{\Gamma} \left| e \rangle_1 \langle e \right|,
\end{equation}
which result into
\begin{equation}
    H_{\text{eff}}= 0, \quad    L_{1,\text{eff}}= i\frac{2\lambda_1}{\sqrt{\Gamma_1}}a^{\dagger} |g\rangle_1\langle g|,
\end{equation}
which gives a linear gain rate $g=4\lambda_1^2/\Gamma_1$ after tracing over the state $|g\rangle_1$. Meanwhile, the constraint conditions for adiabatic elimination give $\Gamma_1\gg\Omega_1\gg g$.

The Hamiltonian to produce the effective nonlinear gain is given by Eq.~(\ref{Eq7}),  which can also be rewritten as $H_{2,1}= H_a+ H_g+H_e +V_-+V_+$ with the ground-state subspace Hamiltonian $ H_g=0$, the excited-state subspace Hamiltonian $H_e=0$, the coupling terms between them $ V_-= \lambda_2 (a^{\dagger})^2 \sigma_{-,2}$ and $ V_+= \lambda_2 a^2 \sigma_{+,2}$, and the Lindblad operators $L = \sqrt{\Gamma} |0 \rangle_2 \langle e|$. Using Eq.~(\ref{HNH}), we can obtain
\begin{equation}
    H_{\text{NH}} = -\frac{i \Gamma_2}{2} \left| e \rangle_2 \langle e \right|, \quad    H_{\text{NH}}^{-1} = \frac{2i}{\Gamma_2}  \left| e \rangle_2 \langle e \right|,
\end{equation}
which result into
\begin{equation}
    H_{\text{eff}}= 0, \quad  L_{2,\text{eff}}= i\frac{2\lambda_2}{\sqrt{\Gamma}}a^{2} |g\rangle_2\langle g|,
\end{equation}
which gives a nonlinear damping rate $\kappa=4\lambda_2^2/\Gamma_2$ after tracing over the ground state $|g\rangle_2$. Meanwhile, the constraint conditions for adiabatic elimination give $\Gamma_2\gg\lambda_2\gg \kappa$.

\section{Classical Van der Pol model}\label{PB}
For our purpose, we reduce the Lindblad master equation in Eq.~(\ref{Eq14}) to the classical equation of the amplitude $\alpha = \langle a \rangle$ in Eq.~(\ref{Eq19}). To achieve this, we can right-multiply both sides of Eq.~(\ref{Eq16}) by the annihilation operator $a$ and then take the trace over it, obtaining
\begin{equation}
\mathrm{Tr}\{\dot{\rho}a\} = -i\mathrm{Tr}\{[H_a , \rho]a\}        +\mathrm{Tr}\{g \mathcal{D}[a^{\dagger}]\rho a\}+\mathrm{Tr}\{\kappa \mathcal{D}[a^{2}]\rho a\}.  \notag
\end{equation}
Substituting $H_a = -\Delta a^{\dagger }a + \varepsilon a^{\dagger} + \varepsilon ^*a$ into it, we obtain
\begin{align}
    -\mathrm{Tr}\{[H_a,\rho]a\} = \Delta \alpha - \varepsilon,
\end{align}
where we have used the cyclic property of the trace. Similarly, the gain and damping terms $\mathrm{Tr}\{ g \mathcal{D}[a^\dagger]\rho a\}$ and $\mathrm{Tr}\{ \lambda \mathcal{D}[a^2]\rho a\}$ can be obtained as
\begin{equation}
\begin{aligned}
\mathrm{Tr}\{g \mathcal{D}[a^{\dagger}]\rho a\} &=\frac{g}{2} \mathrm{Tr}\{\rho a^2a^{\dagger}-\rho aa^{\dagger}a\}= \frac{g}{2} \alpha, \\
\mathrm{Tr}\{\kappa \mathcal{D}[a^{2}]\rho a\} &=\frac{\lambda}{2} \mathrm{Tr}[\rho (a^{\dagger})^2a^3-\rho a(a^{\dagger})^2a^2]= -\lambda |\alpha|^2 \alpha. \notag
\end{aligned}
\end{equation}
Consequently, we arrive at the following equation:
\begin{align}
    \dot{\alpha} = \left(\frac{g-\kappa}{2}+ \mathrm{i}\Delta \right)\alpha - \lambda |\alpha|^2\alpha - \mathrm{i}\varepsilon,
\end{align}
where $\varepsilon \sqrt{\lambda}$ is a constant.

\bibliography{reference}

\end{document}